\documentclass[aps,prl,reprint,showpacs,superscriptaddress]{revtex4-1}
\usepackage[english]{babel}
\usepackage{amsmath,amssymb,bbm,graphicx,color,comment,txfonts}
\usepackage[bookmarks=true,colorlinks,citecolor=blue,urlcolor=blue]{hyperref}
\usepackage{dsfont}

\usepackage{bbm}
\usepackage{float}

\usepackage{dcolumn}   
\usepackage{bm}        
\usepackage{filecontents}
\usepackage{lineno}

\usepackage{mathtools}
\usepackage{xcolor} 

\usepackage[export]{adjustbox}
\usepackage{adjustbox}
\usepackage{subcaption}

\usepackage{braket}

\newcommand\bs[1]{\ensuremath{\boldsymbol{#1}}}

\newcommand{\dagg}{\ensuremath{^{\dagger}}}

\newcommand{\elemA}{\mathord{\vcenter{\hbox{\includegraphics[height=3ex]{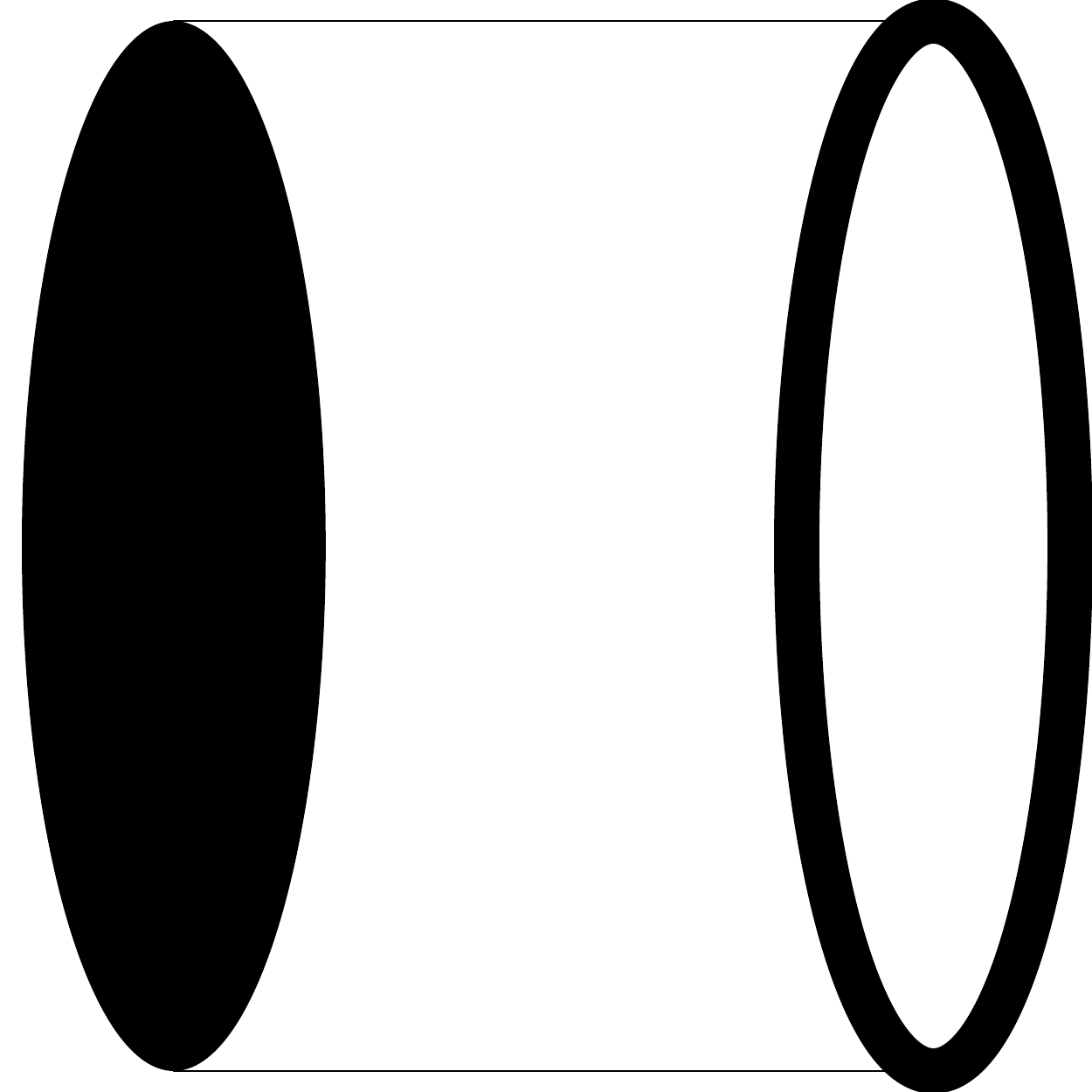}}}}}
\newcommand{\elemB}{\mathord{\vcenter{\hbox{\includegraphics[height=3ex]{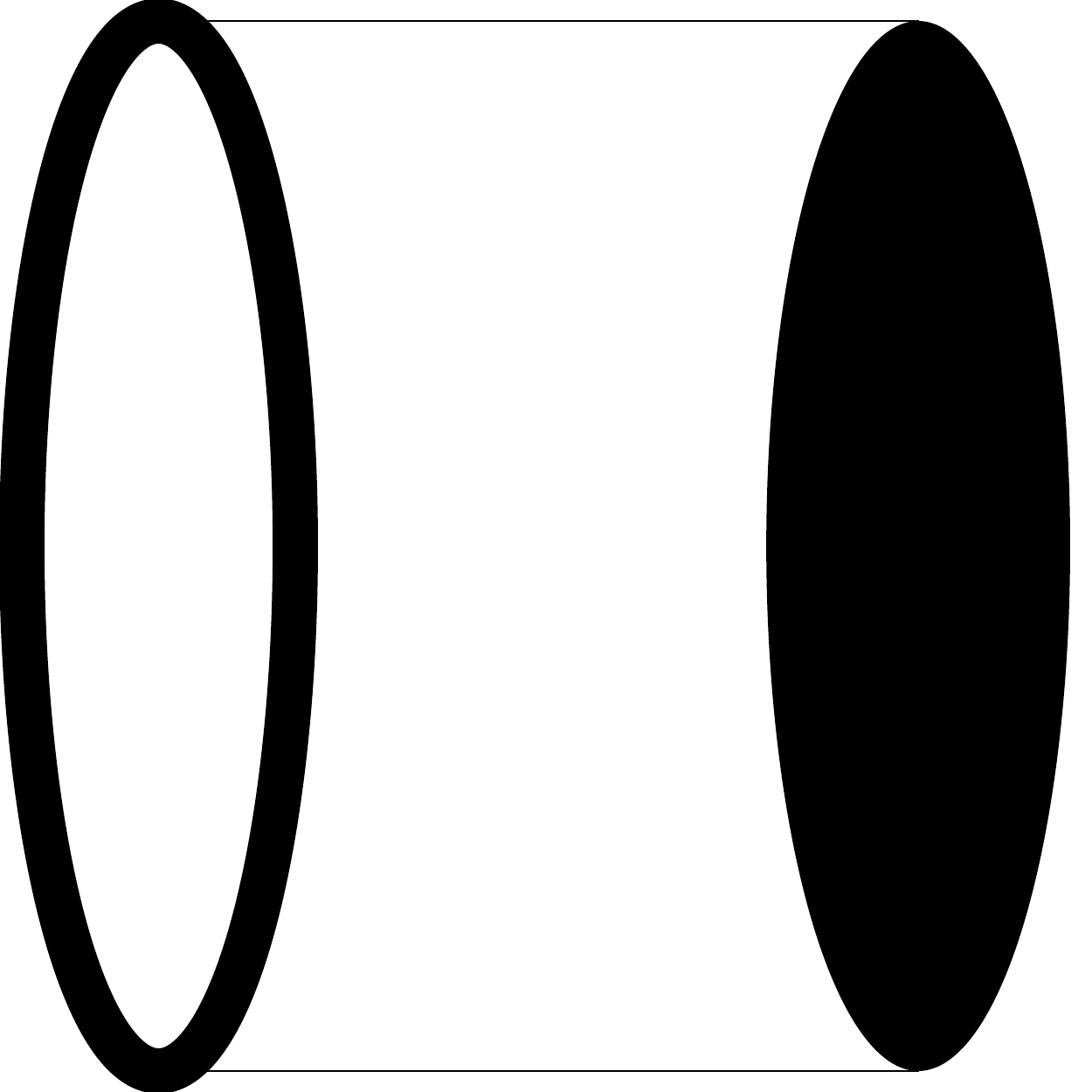}}}}}
\newcommand{\elemC}{\mathord{\vcenter{\hbox{\includegraphics[height=3ex]{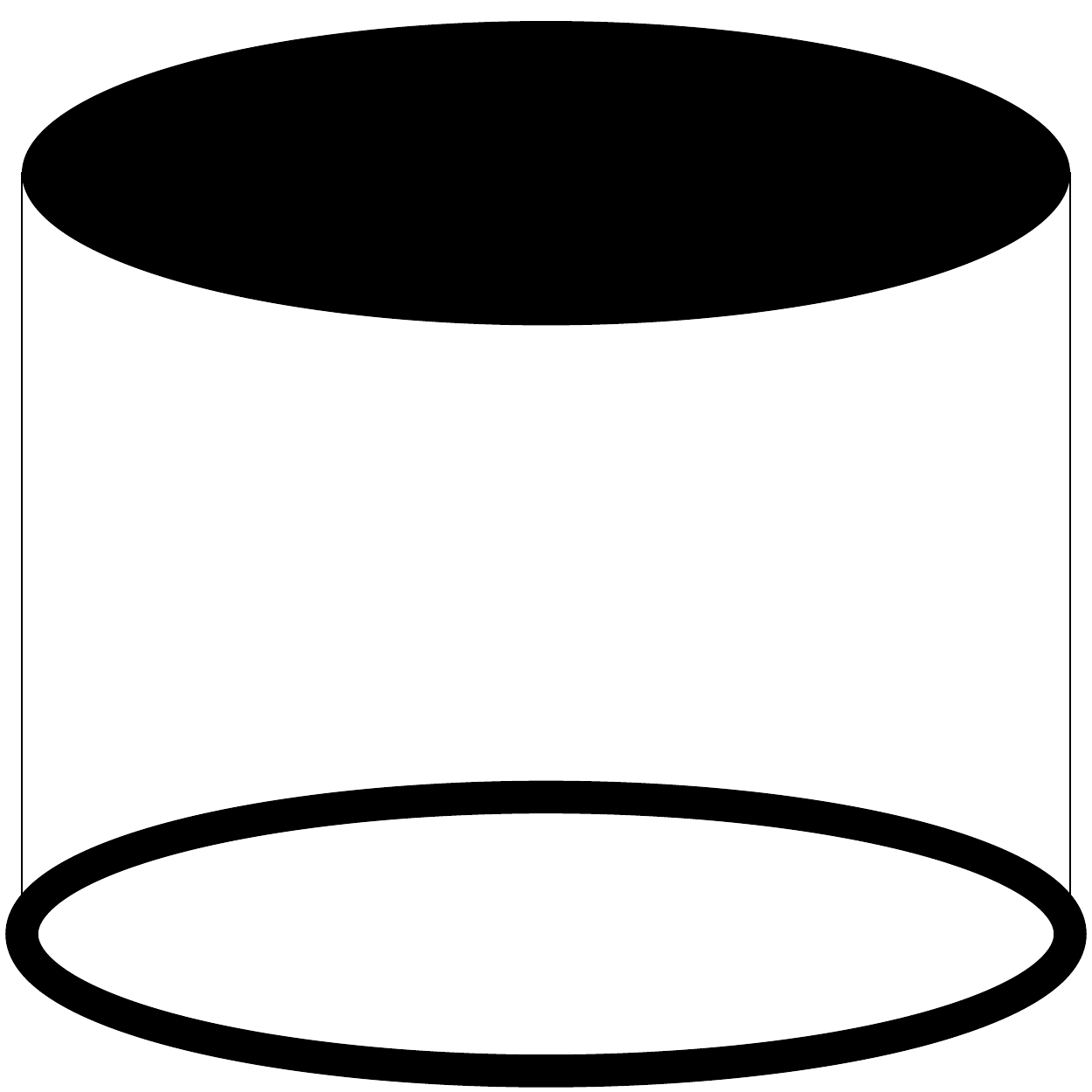}}}}}
\newcommand{\elemD}{\mathord{\vcenter{\hbox{\includegraphics[height=3ex]{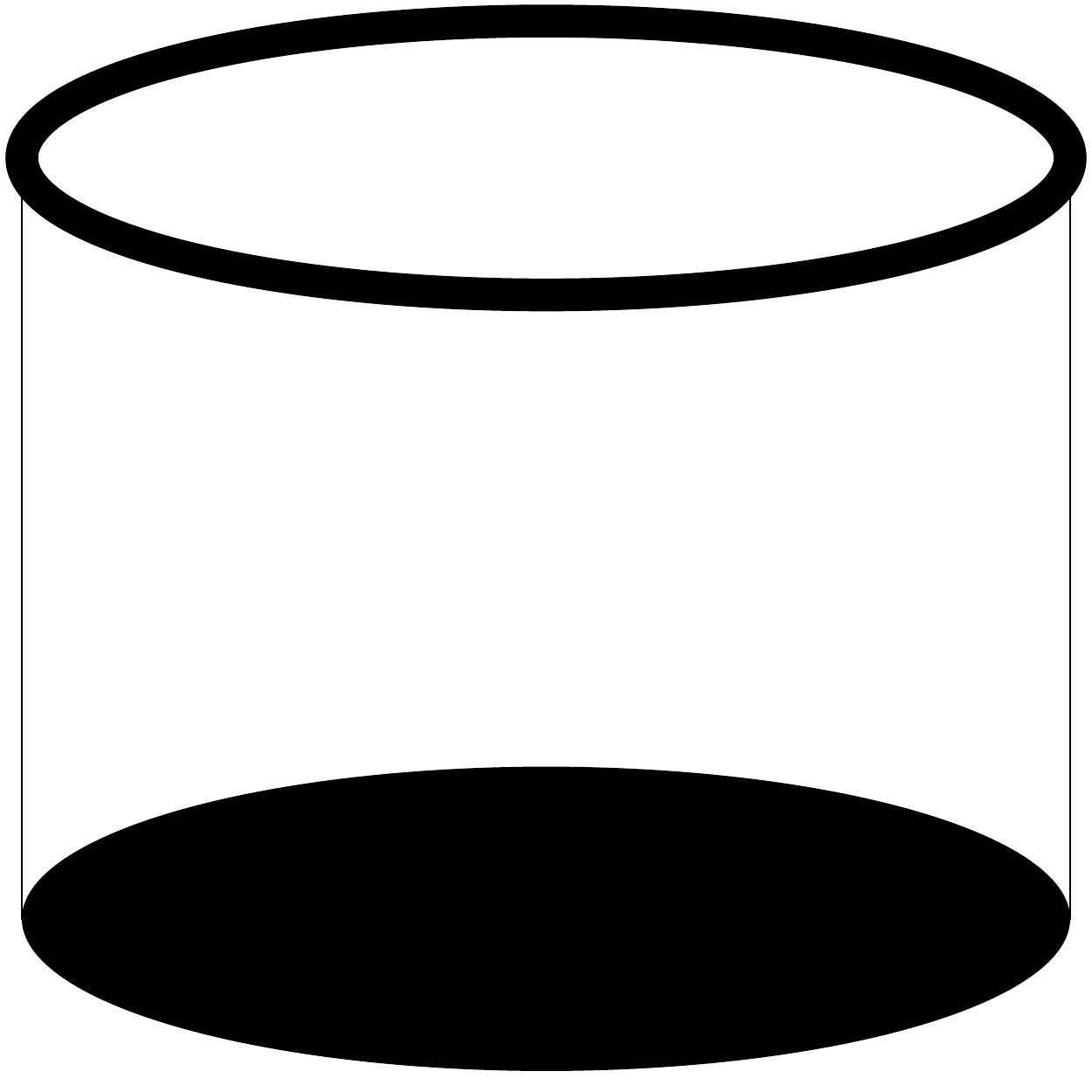}}}}}
\newcommand{\elemE}{\mathord{\vcenter{\hbox{\includegraphics[height=3ex]{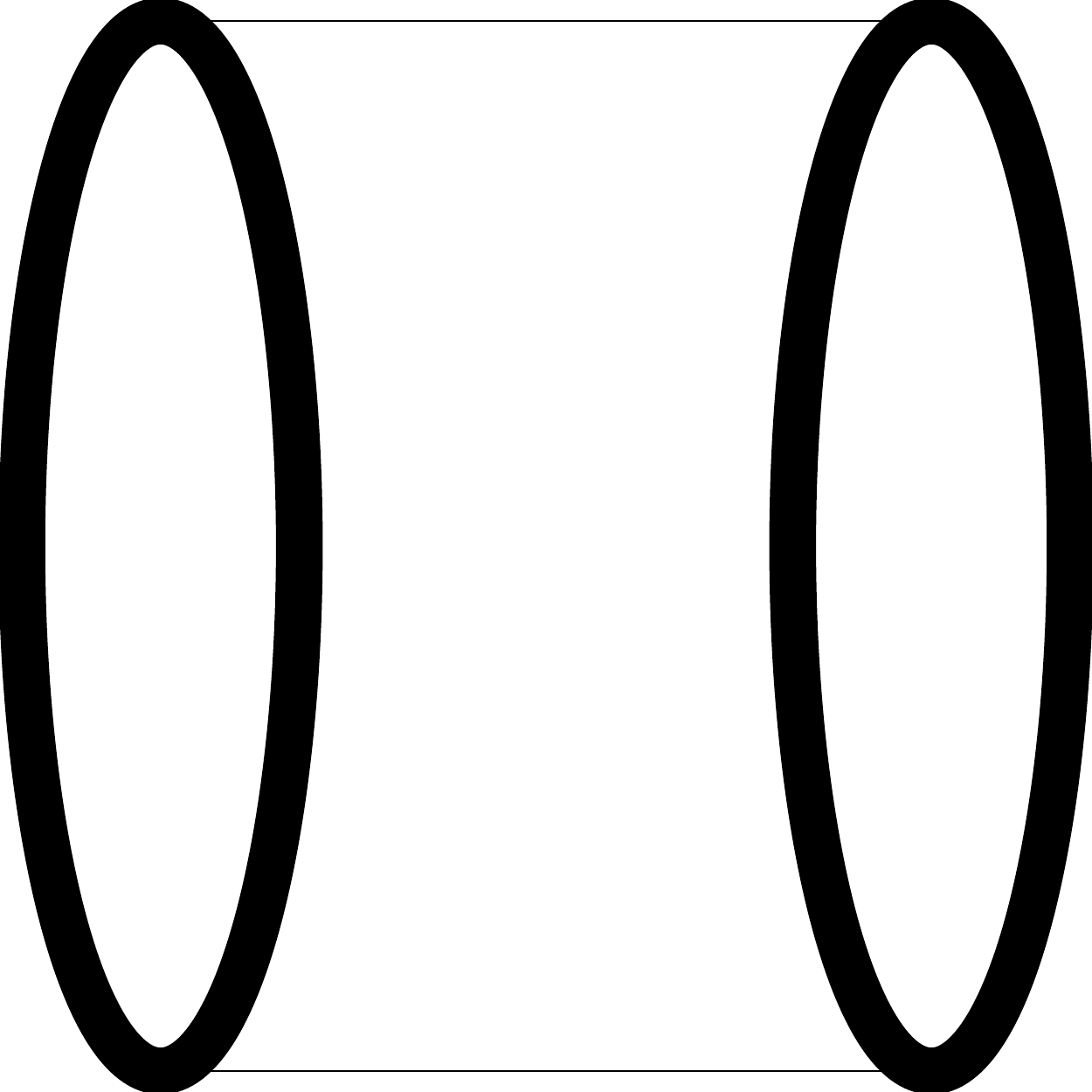}}}}}
\newcommand{\elemF}{\mathord{\vcenter{\hbox{\includegraphics[height=3ex]{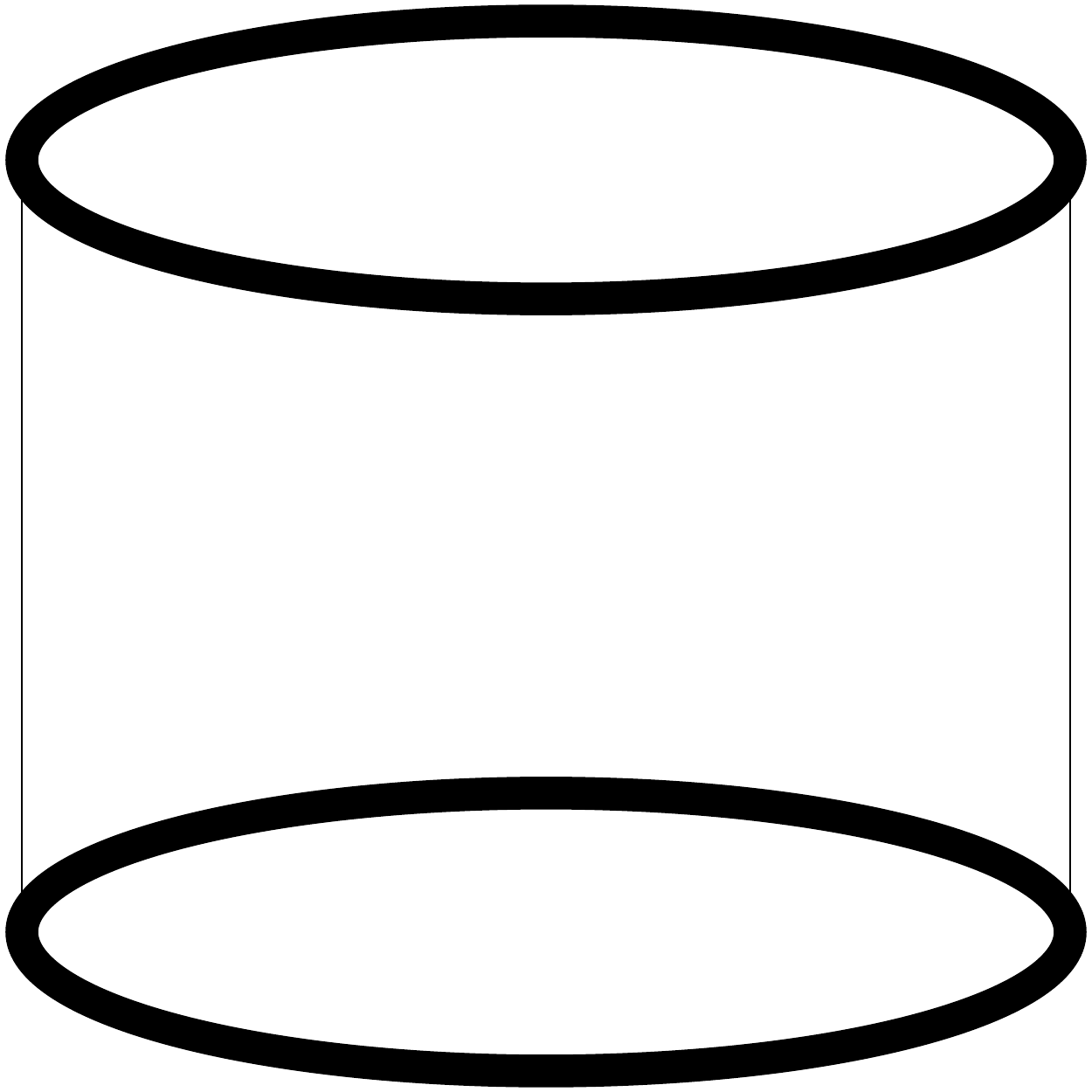}}}}}

\begin{document}

\title{Exact solution of a two-species quantum dimer model for pseudogap metals}
\author{Johannes Feldmeier}
\author{Sebastian Huber}
\author{Matthias Punk}
\date{\today}
\affiliation{Physics Department, Arnold Sommerfeld Center for Theoretical Physics and Center for NanoScience, Ludwig-Maximilians-University Munich, 80333 Munich, Germany}

\begin{abstract}
We present an exact ground state solution of a quantum dimer model introduced in Ref.~\cite{punk2015quantum}, which features ordinary bosonic spin-singlet dimers as well as fermionic dimers that can be viewed as bound states of spinons and holons in a hole-doped resonating valence bond liquid.
Interestingly, this model captures several essential properties of the metallic pseudogap phase in high-$T_c$ cuprate superconductors.
We identify a line in parameter space where the exact ground state wave functions can be constructed at an arbitrary density of fermionic dimers. At this exactly solvable line the ground state has a huge degeneracy, which can be interpreted as a flat band of fermionic excitations. Perturbing around the exactly solvable line, this degeneracy is lifted and the ground state is a fractionalized Fermi liquid with a small pocket Fermi surface in the low doping limit. 
\end{abstract}

\maketitle

Quantum dimer models have been a very useful tool to study paramagnetic ground states of quantum antiferromagnets. Originally introduced by Rokhsar and Kivelson to elucidate the physics of Anderson's resonating valence bond (RVB) state in the context of high-temperature superconductors \cite{Anderson1987, PhysRevB.35.8865, rokhsar1988superconductivity}, these models provide an effective description of low energy singlet excitations in antiferromagnets and feature rich phase diagrams, including a variety of different valence bond solids with broken lattice symmetries, as well as symmetric spin-liquid phases \cite{PhysRevB.40.5204, PhysRevB.54.12938, PhysRevLett.86.1881, PhysRevB.66.214513, PhysRevB.73.245105}. Subsequently, interesting connections to lattice gauge theories and loop gas models have been found, raising interest in quantum dimer models from various perspectives \cite{FradkinKivelsonGauge, PhysRevB.65.024504, Kitaev2003, PhysRevB.69.220403, PhysRevB.83.155117}.

In this work we consider an extension of the Rokhsar-Kivelson (RK) model on the square lattice introduced by Punk, Allais and Sachdev \cite{punk2015quantum}, which provides an effective low-energy description of hole-doped antiferromagnets in two dimensions. The Hilbert space is constructed by hard-core coverings of the square lattice with two flavors of dimers: the standard nearest-neighbor bosonic spin-singlets of the RK model, as well as fermonic dimers carrying charge $+e$ and spin $1/2$. These fermionic dimers can be viewed as bound states of a spinon and a holon in hole-doped RVB states \cite{PhysRevB.39.259, PhysRevB.40.7133, PhysRevB.71.144509, PhysRevB.74.014437, RevModPhys.78.17, PhysRevLett.100.157206, PhysRevB.85.195123}. It has been argued in Refs.~\cite{punk2015quantum, PhysRevB.94.115112} that this model features a so-called fractionalized Fermi liquid ground state \cite{PhysRevLett.90.216403}, with a small Fermi surface enclosing an area proportional to the density of doped holes away from half filling. This apparent violation of Luttinger's theorem, which states that the Fermi surface should enclose an area proportional to the total number of holes with respect to the completely filled band in metallic phases without broken symmetries \cite{PhysRevLett.84.3370}, is possible due to the presence of topological order \cite{PhysRevB.69.035111, doi:10.1093/ptep/ptw110}.

One of the most interesting aspects of this model is the fact that it captures various properties of the metallic pseudogap phase in underdoped high-$T_c$ cuprate superconductors, such as the presence of a small hole-pocket Fermi surface with a highly anisotropic, electronic quasiparticle residue, providing a potential explanation for the observation of Fermi-arcs in photoemission experiments. Moreover, this model exhibits a large pseudogap in the antinodal region of the Brillouin zone around momenta $\bs{k}\sim(0,\pi)$ and symmetry related points \cite{Huber2017}.

While previous studies of this quantum dimer model were mostly based on numerical approaches, we present an exact analytical solution for the ground state at an arbitrary density of fermionic dimers in this work. This solution is based on a generalization of the original idea by Rokhsar and Kivelson that the Hamiltonian can be written as a sum of projectors in certain parameter regimes. While it is easy to see that the corresponding ground state is a simple equal weight superposition of all possible dimer coverings in the RK case, this is no longer true in the presence of fermionic dimers, because the equal weight superposition is not antisymmetric under the exchange of two fermions. Nevertheless, it is still possible to construct the exact ground state wave function, as we show in detail below. Interestingly, we find that fermionic excitations are dispersionless and form a flat band at this generalized RK line in parameter space. Perturbing away from the exactly solvable line we can show that the ground state of this model is indeed a fractionalized Fermi liquid at low densities of fermionic dimers.

We start from the dimer model introduced in Ref.~\cite{punk2015quantum} and add an additional potential energy term for configurations with pairs of parallel fermionic and bosonic dimers within a flippable plaquette configuration. The Hamiltonian $H=H_{RK}+H_1$ consists of two parts: the standard Rokhsar-Kivelson Hamiltonian for bosonic dimers
\begin{equation}
H_{RK} = \sum_{i,\eta} \Big[ - J \, D^\dagger_{i,\eta} D^\dagger_{i+\hat{\bar{\eta}},\eta} D^{\ }_{i, \bar{\eta}} D^{\ }_{i+\hat{\eta}, \bar{\eta}} + V \,  D^\dagger_{i,\eta} D^\dagger_{i+\hat{\bar{\eta}}, \eta} D^{\ }_{i, \eta} D^{\ }_{i+\hat{\bar{\eta}}, \eta} \Big]
\end{equation}
as well as similar terms with plaquette resonances and potential energy terms between a bosonic and a fermionic dimer
\begin{eqnarray}
H_{1} &=& - t_1 \sum_{i} \Big[ D^\dagger_{i,x} F^\dagger_{i+\hat{y},x} F^{\ }_{i,x} D^{\ }_{i+\hat{y},x} + 3 \ \text{terms} \Big] \notag \\
&& + v_1  \sum_{i} \Big[ D^\dagger_{i,x} F^\dagger_{i+\hat{y},x} F^{\ }_{i+\hat{y},x} D^{\ }_{i,x} + 3 \ \text{terms} \Big] \notag \\ 
&& - t_2 \sum_{i} \Big [D^\dagger_{i,x} F^\dagger_{i+\hat{y},x} F^{\ }_{i,y} D^{\ }_{i+\hat{x},y} + 7 \ \text{terms} \Big] \notag \\
&& -t_3 \sum_{i} \Big[D^\dagger_{i,x}F^\dagger_{i+2\hat{x},y}F^{\ }_{i,x}D^{\ }_{i+2\hat{x},y} + 15 \ \text{terms} \Big]
\end{eqnarray}
Here $D_{i,\eta}$ ($F_{i,\eta}$) is an annihilation operator for a bosonic (fermionic) dimer on the bond emanating from lattice site $i$ in direction $\eta \in \{ x,y\}$, while $\hat{\eta} \in \{ \hat{x},\hat{y}\}$ denotes basis vectors in $x$ and $y$ directions (the lattice constant has been set to unity throughout this paper). Finally, $\bar{\eta}$ denotes the complement of $\eta$, i.e.~$\bar{\eta}=x$ if $\eta=y$ and vice versa.
The terms which are not explicitly displayed are related by lattice symmetry operations and hermitian conjugation. Note that in contrast to Ref.~\cite{punk2015quantum} we 
omit a possible spin index for the fermionic dimers. Nevertheless, all our results can be generalized to spinful fermions easily. Further terms involving resonances
of two or more fermionic dimers are possible as well, but are not expected to be important in the interesting regime of low doping, where the density of fermionic dimers is small. Moreover, we will focus exclusively on the topological sector of the Hilbert space of hard-core coverings with zero winding number throughout this work \cite{Huber2017}.

In the next step we identify a line in parameter space which allows us to rewrite the Hamiltonian $H$ as a sum of projectors. As the model then takes a form similar to the original RK-Hamiltonian at $J=V$ \citep{rokhsar1988superconductivity}, we shall speak of an RK line in the following. Setting the parameters to $J=V$, $t_3=0$ and $v_1=t_2=-t_1$, the full Hamiltonian can be expressed graphically as a sum of projectors
\begin{eqnarray}
H &=& J \sum_\text{plaq.} \left( | \elemE \rangle- | \elemF \rangle \right) \left( \langle \elemE | - \langle \elemF |  \right)  + v_1 \sum_{\text{plaq.} \, l} P_l \label{eq:m3}  \\
P_l &=& |\phi_l\rangle \langle \phi_l | \ , \hspace{0.5cm} |\phi_l\rangle =  | \elemA \rangle + | \elemB \rangle - | \elemC \rangle - | \elemD \rangle
\end{eqnarray}
where empty (full) ellipses represent bosonic (fermionic) dimers.

As a consequence of the special form of Eq.~\eqref{eq:m3}, the Hamiltonian is positive definite, i.e.~$\bra{\psi}H\ket{\psi}\geq 0$ for all wave functions $\psi$. The ground state can hence be determined by the condition $H\ket{\psi_0}=E_0\ket{\psi_0}=0$. We now construct ground state wave functions $\ket{\psi_0}$ in an arbitrary sector of the (conserved) number of fermionic dimers $N_f$. In the following calculation we restrict to the case $N_f=2$, the generalization to arbitrary fermion numbers is straightforward. We assume the ground state to be a common eigenstate of $H_{RK}$ and $H_1$. As we already know that the bosonic part $H_{RK}$ is minimized by an equal weight superposition of all hard-core coverings with bosonic dimers, we define the basis states
\begin{equation} \label{eq:egs1}
\begin{aligned}
&\Ket{(i_1,\eta_1),(i_2,\eta_2)}\equiv \\
&=\frac{1}{\sqrt{N_t}}\cdot F\dagg_{i_1,\eta_1}F\dagg_{i_2,\eta_2}\Ket{0}_{(i_1,\eta_1),(i_2,\eta_2)}\otimes\left\{\sum_{c\,\epsilon\,\mathcal{C}_{(i_1,\eta_1),(i_2,\eta_2)}}\Ket{c}\right\},
\end{aligned}
\end{equation}
where the sum runs over all possible bosonic configurations $\Ket{c}$ covering the entire lattice with the exception of the bonds $(i_1,\eta_1)$ and $(i_2,\eta_2)$ which are already occupied by fermionic dimers. Note that $H_\text{RK} \Ket{(i_1,\eta_1),(i_2,\eta_2)} = 0$ is a zero energy eigenstate of $H_{RK}$ by construction. We choose to normalize $\ket{(i_1,\eta_1),(i_2,\eta_2)}$ with respect to the number $N_t$ of all possible classical dimer configurations on the entire lattice. The norm of such a basis state is hence given by
\begin{equation} \label{eq:egs2}
\begin{aligned}
\bigl\Vert\Ket{(i_1,\eta_1),(i_2,\eta_2)}\bigr\Vert^2=\frac{N_{(i_1,\eta_1),(i_2,\eta_2)}}{N_t}=Q_c[(i_1,\eta_1),(i_2,\eta_2)],
\end{aligned}
\end{equation}
where $Q_c[(i_1,\eta_1),(i_2,\eta_2)]$ is the classical dimer correlation function. $N_{(i_1,\eta_1),(i_2,\eta_2)}$ denotes the number of all classical configurations with two dimers fixed at $(i_1,\eta_1)$ and $(i_2,\eta_2)$. With these correlations we implicitly enforce the hard-core constraint, as any constraint-violating configuration $C$ yields a vanishing norm $Q_c[C]=0$.

In order to construct a ground state $|\psi_0 \rangle$ of the full Hamiltonian $H=H_{RK}+H_1$ we start with a general expansion
\begin{equation} \label{eq:egs3}
\Ket{\psi_0}=\sum_{i_1,\eta_1,i_2,\eta_2}A_{(i_1,\eta_1),(i_2,\eta_2)}\Ket{(i_1,\eta_1),(i_2,\eta_2)} \ .
\end{equation}
Applying the Hamiltonian we obtain
\begin{equation} \label{eq:egs4}
H\Ket{\psi_0}=v_1 \sum_l\sum_{i_1\eta_1,i_2,\eta_2}A_{(i_1,\eta_1),(i_2,\eta_2)}\;P_l\Ket{(i_1,\eta_1),(i_2,\eta_2)} \ .
\end{equation}
Note that $P_l$ acts nontrivially only on plaquettes containing a single fermionic dimer and thus 
\begin{equation} \label{eq:egs5}
\begin{aligned}
P_l&\Ket{(i_1,\eta_1),(i_2,\eta_2)}=\\
&=\left(\delta_{l,i_1}+\delta_{l+\hat{\bar{\eta}}_1,i_1}+\delta_{l,i_2}+\delta_{l+\hat{\bar{\eta}}_2,i_2}\right)P_l\Ket{(i_1,\eta_1),(i_2,\eta_2)}.
\end{aligned}
\end{equation}
Furthermore, we find
\begin{equation} \label{eq:egs6}
\begin{aligned}
\delta_{l,i_1}P_l\Ket{(i_1,\eta_1),(i_2,\eta_2)}=\delta_{l,i_1}\left(-1\right)^{s_{\eta_1}}\Ket{\phi_{l},(i_2,\eta_2)}
\end{aligned}
\end{equation}
and similar relations for the remaining three terms of Eq.~\eqref{eq:egs5}, where we defined the states
\begin{equation} \label{eq:egs7}
\Ket{\phi_l,(i,\eta)}=\frac{1}{\sqrt{N_t}}F\dagg_{i,\eta}\Ket{0}_{(i,\eta)}\otimes\Ket{\phi_l}\otimes\left\{\sum_{c\,\epsilon\,\mathcal{C}_{(l,x),(l+\hat{y},x),(i,\eta)}} \Ket{c}\right\},
\end{equation}
and further $s_{\eta=x}=1$, $s_{\eta=y}=0$. Again, normalization of these states resorts to classical correlations and effectively projects onto the physical space of hard-core configurations.

Inserting Eq.~\eqref{eq:egs6} into \eqref{eq:egs5} and \eqref{eq:egs4}, and demanding that all coefficients for the states $\Ket{\phi_l,(i_2,\eta_2)}$ vanish results in the two conditions
\begin{equation} \label{eq:egs8}
\begin{aligned}
&A_{(l,x),(i_2,\eta_2)}-A_{(l,y),(i_2,\eta_2)}+A_{(l+\hat{y},x),(i_2,\eta_2)}-A_{(l+\hat{x},y),(i_2,\eta_2)}=0 \\
&A_{(i_1,\eta_1),(l,x)}-A_{(i_1,\eta_1),(l,y)}+A_{(i_1,\eta_1),(l+\hat{y},x)}-A_{(i_1,\eta_1),(l+\hat{x},y)}=0 \ ,
\end{aligned}
\end{equation}
which can be solved by a simple product Ansatz
$A_{(i_1,\eta_1),(i_2,\eta_2)}=a_{i_1,\eta_1}a_{i_2,\eta_2}$,
leading to
\begin{equation} \label{eq:egs10}
a_{i_m,x}-a_{i_m,y}+a_{i_m+\hat{y},x}-a_{i_m+\hat{x},y}=0
\end{equation}
for $m=1,2$. At this point, the generalization to an arbitrary number of fermionic dimers in the system is straightforward and can be done by extending Eq.~\eqref{eq:egs10} to $m=1,...,N_f$. We introduce the lattice momenta $\bs{p}_m$ and make the Ansatz
\begin{equation} \label{eq:egs11}
a_{i_m,\eta_m} = a_{i_m,\eta_m}(\bs{p}_m)=C_{\eta_m}(\bs{p}_m)e^{i\bs{p}_m\cdot \bs{i}_m},
\end{equation}
where the factors $C_{\eta}(\bs{p})$ can be interpreted as weight factors for the two possible dimer orientations and $\bs{i}_m$ denotes the lattice position of site $i_m$. Using this Ansatz in Eq.~\eqref{eq:egs10} and choosing the normalization $|C_x(\bs{p})|^2+|C_y(\bs{p})|^2=\frac{4}{N}$ for later convenience, we obtain
\begin{equation} \label{eq:egs12}
\begin{aligned}
C_{\eta}(\bs{p})=&\frac{2}{\sqrt{N}}\frac{1+e^{ip_{\eta}}}{\sqrt{\left|1+e^{ip_y}\right|^2+\left|1+e^{ip_x}\right|^2}},
\end{aligned}
\end{equation}
where $N$ is the number of lattice sites. One can thus write exact ground states of $H$ on the RK line with two fermionic dimers as
\begin{equation} \label{eq:egs13}
\Ket{\psi_0}=\Ket{\bs{p}_1,\bs{p}_2}=\sum_{i_1,\eta_1,i_2,\eta_2}a_{i_1,\eta_1}(\bs{p}_1)a_{i_2,\eta_2}(\bs{p}_2)\Ket{(i_1,\eta_1),(i_2,\eta_2)}.
\end{equation}
Note that $\bs{p}_1$ and $\bs{p}_2$ take arbitrary values in the first Brillouin zone and $\Ket{\bs{p}_1,\bs{p}_2}=-\Ket{\bs{p}_2,\bs{p}_1}$ is antisymmetric under the exchange of $\bs{p}_1$ and $\bs{p}_2$. The ground state degeneracy corresponds to the $N(N-1)/2$ possibilities to choose $\bs{p}_1, \bs{p}_2$. Interestingly, this result implies that fermionic dimers have a flat dispersion at the RK-line, which we confirmed independently by an exact diagonalization of the Hamiltonian on the RK line for a finite system. We also note that the state in Eq.~\eqref{eq:egs13} is properly normalized in the limit $N\rightarrow \infty$.

For an arbitrary number $N_f$ of fermionic dimers the ground states take the form $\Ket{\psi_0}=\Ket{\bs{p}_1,...,\bs{p}_{N_f}}$ and there are $N!/((N-N_f)!N_f!)$ possibilities to choose the $N_f$ momenta $(\bs{p}_1,...,\bs{p}_{N_f})$.  It is important to emphasize that the states $\Ket{\bs{p}_1,...,\bs{p}_{N_f}}$ are in general \emph{not} linearly independent, and the number of possible momenta $(\bs{p}_1,...,\bs{p}_{N_f})$ does not correspond to the ground state degeneracy in sectors with a large density of fermionic dimers. In fact, it is easy to see that the number of possible choices for the $N_f$ momenta exceeds the number of basis states at large $N_f$. However, in the low doping limit
\begin{equation} \label{eq:limit}
N_f=const.,\;  N \to \infty, 
\end{equation}
 the $\Ket{\bs{p}_1,...,\bs{p}_{N_f}}$ become orthonormal and we indeed obtain the ground state degeneracy via the above relation. 
 
It is instructive to note how the states  $\Ket{\bs{p}}$ for $N_f=1$ are related to the usual bosonic RK ground state, if the fermionic dimer is replaced with a bosonic one. 
As shown in the supplementary material, the purely bosonic states $\Ket{\bs{p}}$ vanish identically for $\bs{p}\neq 0$, which only leaves the ordinary RK-state with $\bs{p}=0$, i.e.~the equal superposition of all bosonic dimer coverings, as the unique ground state.

In the following we want to study how perturbations $\Delta H$ of the Hamiltonian away from the RK line change the ground state structure. We consider perturbations of the form $H+\Delta H = H(t_i \to t_i +\delta t_i)$. As expected, the huge ground-state degeneracy will be lifted and the fermions will acquire a dispersion. The perturbative ground-state in the vicinity of the RK line is then unique and similar to a Fermi gas, where the lowest energy momentum states $\bs{p}_m$ will be filled with $N_f$ fermions. We restrict our discussion to the limit of Eq.~\eqref{eq:limit}, where the degenerate ground states $| \bs{p}_1,\dots \bs{p}_{N_f} \rangle$ are properly normalized. Moreover, we only consider terms in $\Delta H$ which exchange two dimers, i.e.~$\delta t_1$ and $ \delta t_3$ terms. Flip interactions like $t_2$ will be neglected for simplicity, but can be included as well. 

\begin{figure*}[] 
\begin{center}
\includegraphics[trim={0 3cm 0 3cm},clip,width=0.98 \textwidth]{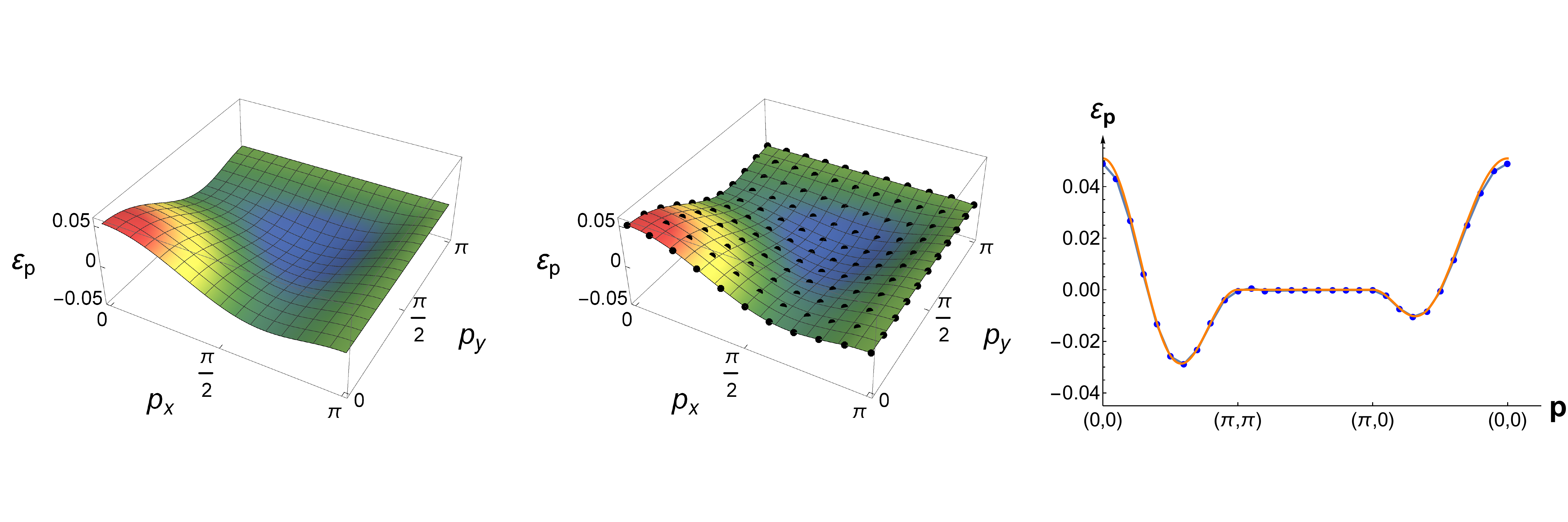}
\caption{Comparison between $\varepsilon (\bs{p})$ from Eq.~\eqref{eq:pert3} (left) and the dispersion obtained from exact diagonalization (ED) for $6\times6$ lattice sites with one fermionic dimer and twisted boundary conditions (middle) for $J=V=1$, $v_1=t_2=-t_1=1$ and $\delta t_3=-0.02$. Right: corresponding line cut through the Brillouin zone (blue line with dots: ED, orange line: Eq.~\eqref{eq:pert3}). }
\label{fig:fRK1}
\end{center}
\end{figure*}

Within first order perturbation theory the eigenstates remain unchanged, but their energy is given by $\Delta E = \langle \bs{p}_1,\dots \bs{p}_{N_f} | \Delta H | \bs{p}_1,\dots \bs{p}_{N_f} \rangle$.
Evaluating the matrix elements for the case $N_f=2$ we get $\Delta E = \varepsilon(\bs{p}_1)+\varepsilon(\bs{p}_2)$ with 
\begin{equation} \label{eq:pert3}
\begin{aligned}
\varepsilon(\bs{p})&=-4\sum_{i=1,3}\delta t_i\, Q_c[(0,x),(r^{1,x}_{t_i},x+\eta_{t_i})]\times \\
&\times\sum_{\eta}\frac{(1+e^{ip_{\eta}})(1+e^{-ip_{\eta+\eta_{t_i}}})}{\left|1+e^{ip_{y}}\right|^2+\left|1+e^{ip_{x}}\right|^2}\sum_{s=1}^{S_{t_i}}\left[e^{-ir^{s,\eta}_{t_i}\cdot \bs{p}}\right],
\end{aligned}
\end{equation}
where $r_{t_i}^{s,\eta}$ and $\eta_{t_i}$ correspond to displacement vector and relative change in orientation for a given $t_i$ process which annihilates a fermionic dimer with initial orientation $\eta$. The sum over the possible $r_{t_i}^{s,\eta}$ corresponding to a given $t_i$ depends on the orientation index $\eta$ and runs from $s=1$ to $S_{t_1}=2$, $S_{t_3}=8$. The classical probabilities $Q_c[(0,x),(r^{1,x}_{t_i},x+\eta_{t_i})]$ are $1/8$ and $1/(4\pi)$ for $t_1$ and $t_3$ respectively and can be obtained from the exact solution of the classical dimer problem \cite{fisher1963statistical, samuel1980use}.
Details of the computation can be found in the supplementary material.
We show an example for $\varepsilon (\bs{p})$ together with exact diagonalization results on a $6\times 6$ lattice with one fermionic dimer and twisted boundary conditions in Fig.~\ref{fig:fRK1}. For $|\delta t_i| \ll |v_1|, J$ we find excellent agreement. Note the formation of hole-pockets around $(\pi/2,\pi/2)$ at a finite density of fermionic dimers for perturbations in $\delta t_3$.

The preceeding results demonstrate that the energy of a state $\Ket{\bs{p}_1,...,\bs{p}_{N_f}}$ is additive in the single particle energies in the low doping limit, indicating a system with Fermi-liquid like behaviour. Now we show that in the same limit the ground states $\Ket{\bs{p}_1,...,\bs{p}_{N_f}}$ can be constructed using creation and annihilation operators that fulfill canonical fermionic anticommutation relations.

We start by defining the vaccum state of the theory to be the usual  RK-ground state, i.e. $\Ket{0^*}=\Ket{\text{RK}}$, which corresponds to the equal weight superposition of all possible hard-core coverings of the lattice with bosonic dimers. We add the star in this notation to emphasize the difference to the vacuum state $\Ket{0}$ used previously. By defining the operator
\begin{equation} \label{eq:FL1}
f\dagg_{\bs{p}}=\sum_{i,\eta}a_{i,\eta}(\bs{p})F\dagg_{i,\eta}D^{\ }_{i,\eta},
\end{equation}
we can express the possible ground states along the RK-line as
\begin{equation} \label{eq:FL2}
\Ket{\bs{p}_1,...,\bs{p}_{N_f}}=\prod_{i=1}^{N_f}f\dagg_{\bs{p}_i}\Ket{0^*}.
\end{equation}
We aim to show that the corresponding Hamiltonian $H=\sum_{\bs{p}}\varepsilon (\bs{p})f\dagg_{\bs{p}}f_{\bs{p}}$ describes the model in the vicinity of the RK-line as a system of non-interacting fermionic excitations. We hence need to show that the canonical anticommutation relations
\begin{equation} \label{eq:FL3}
\{f\dagg_{\bs{p}_1},f_{\bs{p}_2}\}=\delta_{\bs{p}_1,\bs{p}_2}
\end{equation}
are satisfied in the limit of Eq.~\eqref{eq:limit}. Note that we require specification of the Hilbert space on which Eq.~\eqref{eq:FL3} is supposed to hold. In usual fermionic theories the anticommutation relations must hold on the Fock space spanned by the set of states $\{\prod_{i=1}^{N_f}c\dagg_{\bs{k}_i}\Ket{0}\}$. In direct analogy we demand that in our model Eq.~\eqref{eq:FL3} should hold on the Hilbert space spanned by the states $\{\prod_{i=1}^{N_f}f\dagg_{\bs{k}_i}\Ket{0^*}\}$. Thus, even though the operators of Eq.~\eqref{eq:FL1} do not constitute fermionic operators on a Hilbert space built upon the actual vacuum state $\Ket{0}$, we still can prove them to be fermionic within our relevant Hilbert space. The quantity we aim to compute is now $\{f\dagg_{\bs{p}_1},f_{\bs{p}_2}\}\Ket{0^*}$ and we want to show that this expression yields $\delta_{\bs{p}_1,\bs{p}_2}\Ket{0^*}$. From the relation $\{f\dagg_{\bs{p}_1},f_{\bs{p}_2}\}=\sum_{i,\eta}a_{i,\eta}(\bs{p}_1)a^*_{i,\eta}(\bs{p}_2)\,\hat{N}_{i,\eta}$, where $\hat{N}_{i,\eta}$ corresponds to the total dimer number operator on the link $(i,\eta)$, we deduce that $\Vert\{f\dagg_{\bs{p}_1},f_{\bs{p}_2}\}\Ket{0^*}\Vert^2$ is given by the Fourier transformed classical dimer correlation function (see supplementary material for details), which reduces to $\delta_{\bs{p}_1,\bs{p}_2}$ for $N \to \infty$ as claimed, with corrections of order $\mathcal{O}(\log (N)/N)$. The appearance of the total dimer number operator $\hat{N}_{i,\eta}$ then ensures that this result remains valid for all states in our Hilbert space provided that Eq.~\eqref{eq:limit} be fulfilled. Beyond this limit, where the Fourier transform of the classical dimer correlation function reduces to a delta function, we show in the supplementary material that Eq.~\eqref{eq:FL3} is exact for arbitrary system sizes if the momenta $\bs{p}_1$ and $\bs{p}_2$ lie on the Brillouin zone diagonal.

Finally we can also relate the operator $f_{\bs{p}}$ to the actual electron annihilation operator $c_{\bs{p}}$. In Ref.~\cite{punk2015quantum} it was shown that the electron annihilation operator in the dimer Hilbert space takes the form
\begin{equation}
c_{\bs{p},\alpha} = \frac{\varepsilon_{\alpha \beta}}{2 \sqrt{N}} \sum_{j,\eta} \left(1+e^{-i p_\eta}\right) F^\dagger_{i,\eta,\beta}D^{\ }_{i, \eta} e^{-i \bs{p} \cdot \bs{i}_j} \ ,
\end{equation}
i.e.~removing an electron on lattice site $i$ corresponds to replacing a bosonic with a fermionic dimer on all adjacent bonds. Here we included a spin index $\alpha$ and $\varepsilon_{\alpha\beta}$ is the unit antisymmetric tensor. Surpressing the electronic spin index and comparing this expression with the definition of the $f_{\bs{p}}$ operator in Eq.~\eqref{eq:FL1} it immediately follows that
\begin{equation} \label{eq:FL4}
f\dagg_{\bs{p}}=\frac{4}{\sqrt{|1+e^{ip_x}|^2+|1+e^{ip_y}|^2}}\,c_{-\bs{p}} \ .
\end{equation}
This relation is particularily useful, because it shows that the Fermi surface of fermionic dimers directly translates to the electronic Fermi surface. 
Moreover, from the fact that the $f_{\bs{p}}$ fermions form a free Fermi gas, we can infer that the electron spectral function in the vicinity of the RK-line takes the form
$A_e(\bs{p},\omega) = \mathcal{Z}_{\bs{p}} \, \delta(\omega-\varepsilon(\bs{p}))$
with a quasiparticle weight $ \mathcal{Z}_{\bs{p}} = \big[\cos^2(p_x/2) + \cos^2(p_y/2)\big]/4$, which is distributed anisotropically around the Fermi surface (see also Refs.~\cite{punk2015quantum, Huber2017} for a discussion. Similar results have been obtained in a SU(2) slave-particle approach \cite{Bieri2009} as well as in projected wave function studies \cite{Bieri2007}). Note that the electron spectral function at the RK line only features a coherent peak, but no incoherent background. Perturbing away from the RK line incoherent weight appears, but not within first order perturbation theory. Numerical results obtained by ED confirm this result.

In summary, we provided an exact ground state solution for the dimer model introduced in Ref.~\cite{punk2015quantum} on a particular line in parameter space, for arbitrary densities of fermionic dimers. At this line the ground state is massively degenerate and can be interpreted as a fermionic flat band. Perturbing away from the exactly solvable line lifts this degeneracy and we were able to show that the ground state is a fractionalized Fermi liquid, at least in the limit of small fermionic dimer densities. In this limit the ground state can be constructed by applying canonical fermion creation operators to a suitably choosen vacuum state and the energy of these fermions is additive. Moreover, these fermionic operators are directly related to electron creation operators in the restricted Hilbert space of our model. Even though we limited the discussion to spinless fermionic dimers, our construction can be easily generalized to spin-1/2 fermionic dimers. We also note that the very same construction works for other lattice geometries as well, such as a triangular lattice, where we expect that the fractionalized Fermi liquid ground state is stable over a wider parameter regime. Indeed, the $U(1)$ spin liquid in the square lattice RK model at half filling is unstable towards confining, symmetry broken states away from the special RK point $J=V$. On non-bipartite lattices an extended $Z_2$ spin liquid phase exists, however \cite{PhysRevLett.86.1881}. Analogous considerations hold for hole doped RK models \cite{Ribeiro2007, Lamas2012} as well as the fractionalized Fermi liquid phase discussed here \cite{doi:10.1093/ptep/ptw110, PhysRevB.93.165139}. Including diagonal, next-nearest neighbour dimers in our model is thus an interesting point for future study. In conclusion, our results provide a rare example of a strongly correlated, fermionic lattice model in two dimensions, which is exactly solvable and potentially relevant for the description of the metallic pseudogap phase in underdoped cuprates.

\acknowledgements

This research was supported by the German Excellence Initiative via the Nanosystems Initiative Munich (NIM).

\bibliography{RKlit}
\bibliographystyle{ieeetr}

\section{Supplemental Material}



\subsection{Matrix elements} \label{matrixelements}
We show how to evaluate matrix elements in order to compute the normalization of the states $\Ket{\bs{p}_1,...,\bs{p}_{N_f}}$ as well as the first order perturbative ground state energy $\Delta E=\Braket{\bs{p}_1,...,\bs{p}_{N_f}|\Delta H|\bs{p}_1,...,\bs{p}_{N_f}}$. To abbreviate the arising expressions we employ the notation $i_1=(i_1,\eta_1)$ for denoting the links of the lattice. The overlap of two basis states defined in Eq.(\ref{eq:egs1}) from the main text is given by
\begin{equation} \label{eq:me1}
\Braket{l_1,...,l_{N_f}|i_1,...,i_{N_f}}\propto \sum_{P\,\in\,\pi_{N_f}}(-1)^{\sigma(P)}\delta_{i_1,l_{P(1)}}\cdot ... \cdot \delta_{i_{N_f},l_{P(N_f)}},
\end{equation}
where the sum runs over all permutations $P$ of $1,...,N_f$ and the proportionality constant will be given by the classical dimer correlation function $Q_c[i_1,...,i_{N_f}]$. The sign $(-1)^{\sigma(P)}$ is $-1$ for odd permutations, $+1$ for even permutations and is due to exchanging the anticommuting fermionic operators $F$, $F^\dagger$.
The matrix elements arising in the evaluation of the first order perturbative ground state energy $\Delta E$ for general fermion numbers $N_f$ can then be evaluated as
\begin{widetext}
\begin{equation} \label{eq:me3}
\Braket{l_1,...,l_{N_f}|F\dagg_{j+r}D\dagg_jD_{j+r}F_j|i_1,...,i_{N_f}}=\sum_{k=1}^{N_f}\delta_{j,i_k}\sum_{P\,\in\,\pi_{N_f}}(-1)^{\sigma(P)}\delta_{i_1,l_{P(1)}}\cdot ... \cdot \delta_{i_k+r,l_{P(k)}}\cdot ... \cdot \delta_{i_{N_f},l_{P(N_f)}}\;Q_c[i_1,....i_k,i_k+r,...,i_{N_f}].
\end{equation}
\end{widetext}

\subsection{Normalization}
Using Eq.~\eqref{eq:me1} and \eqref{eq:me3}, ground state normalizations and the perturbative energy change $\Delta E$ can now be evaluated. We start by considering the case $N_f=2$ and compute the norm of the ground state from Eq.(\ref{eq:egs13}) from the main text via Eq.~\eqref{eq:me1},
\begin{equation} \label{eq:me4}
\begin{aligned}
&\Braket{\bs{p}_1,\bs{p}_2|\bs{p}_1,\bs{p}_2}=\sum_{i_1,\eta_1,i_2,\eta_2}\biggl\{\left|a_{i_1,\eta_1}(\bs{p}_1)\right|^2\left|a_{i_2,\eta_2}(\bs{p}_2)\right|^2-\\
&-a_{i_1,\eta_1}(\bs{p}_1)a^*_{i_1,\eta_1}(\bs{p}_2)a_{i_2,\eta_2}(\bs{p}_2)a^*_{i_2,\eta_2}(\bs{p}_1)\biggr\}Q_c\left[(i_1,\eta_1),(i_2,\eta_2)\right].
\end{aligned}
\end{equation}
Upon inserting the expression for $a_{i,\eta}(\bs{p})$ from the main text, we find that the second term in Eq.~\eqref{eq:me4} contains the Fourier transform of the classical dimer correlation function. The classical dimer model is an exactly solvable problem that can be treated using Pfaffians. We resort to the well known long-range properties of these classical correlations that were analyzed in Refs.~\cite{fisher1963statistical, youngblood1980},
\begin{equation} \label{eq:me5}
\begin{aligned}
Q_c[(0,x),&((X,Y),x)]=\\
&=\frac{1}{16}+\frac{1}{2\pi^2}\left[(-1)^{X+Y}\frac{Y^2-X^2}{R^4}+(-1)^X\frac{1}{R^2}\right],
\end{aligned}
\end{equation}
where $R^2=X^2+Y^2$. The algebraic decay of these correlations then ensures that in a large system, the Fourier transformed classical dimer correlations take the form
\begin{equation} \label{eq:me6}
\frac{1}{N^2}\sum_{i_1,i_2}e^{i(\bs{i}_1-\bs{i}_2)\cdot \bs{p}}Q_c[i_1,i_2]\rightarrow \frac{1}{4}\delta_{\bs{p},0}+\mathcal{O}\left(\log(N)/N\right),
\end{equation}
i.e.~a Kronecker delta with corrections of order $\log(N)/N$, depending on the chosen momentum point $\bs{p}$. As $\Ket{\bs{p}_1,\bs{p}_2}$ is antisymmetric, the norm of Eq.~\eqref{eq:me4} must vanish for $\bs{p}_1=\bs{p}_2$. Therefore, for $\bs{p}_1\neq \bs{p}_2$ the second term in Eq.~\eqref{eq:me4} only contains the Fourier transform of $Q_c$ at some finite momentum and hence vanishes for $N\rightarrow \infty$. Evaluating the first term and again neglecting corrections of maximum order $\log(N)/N$ we find $\Braket{\bs{p}_1,\bs{p}_2|\bs{p}_1,\bs{p}_2}=1$. 

This normalization generalizes to arbitrary $N_f\ll N$ via the use of Eq.~\eqref{eq:me1}. Evaluating $\Braket{\bs{p}_1,...,\bs{p}_{N_f}|\bs{p}_1,...,\bs{p}_{N_f}}$, every permutation in Eq.~\eqref{eq:me1} other than the trivial permutation will involve a finite momentum Fourier transform of the classical dimer correlation function of $N_f$ dimers. As the (long-range) dimer correlations can be constructed from a quadratic field theory using Grassmann variables \cite{samuel1980use}, Wick's theorem ensures that finite momentum Fourier transforms of higher order correlation functions still contribute corrections of maximum order $\mathcal{O}(\log(N)/N)$ to the ground state normalization. The remaining trivial permutation then yields normalization to unity, again dropping all Fourier transformed classical correlations. Note that within the limit of Eq.(\ref{eq:limit}) from the main text, the states $\Ket{\bs{p}_1,...,\bs{p}_{N_f}}$ form an orthonormal set, where orthogonality can be shown by the same reasoning as normalization, as for $(\bs{p}_1,...,\bs{p}_{N_f})\neq (\bs{q}_1,...,\bs{q}_{N_f})$ all terms arising from the evaluation of the matrix elements will now contain a finite momentum Fourier transform of $Q_c$.

\subsection{Perturbation theory}
We briefly discuss how the dispersion displayed in Eq.(\ref{eq:pert3}) from the main text can be obtained using Eq.~\eqref{eq:me3} for the matrix elements. The starting point is a perturbed Hamiltonian relative to the RK-line, with perturbation in arbitrary exchange interactions:
\begin{equation} \label{eq:pt1}
\Delta H=\delta t_i\sum_{s=1}^{S_{t_i}}\sum_{j,\eta}F\dagg_{j+r_{t_i}^{s,\eta},\eta+\eta_{t_i}}D\dagg_{j,\eta}D_{j+r_{t_i}^{s,\eta},\eta+\eta_{t_i}}F_{j,\eta}.
\end{equation}
As already mentioned in the main text, $r^{s,\eta}_{t_i}$ with $s\in\{1,...,S_{t_i}\}$ labels the possible displacement vectors of a $t_i$-process which annihilates a fermionic dimer with orientation $\eta$, $\eta_{t_i}$ is the change in dimer orientation in the course of the process. The expectation value of $\Delta H$ with respect to the $N_f=2$ ground states $\Ket{\bs{p}_1,\bs{p}_2}$ can then be expressed as
\begin{widetext}
\begin{equation} \label{eq:pt2}
\begin{aligned}
\Delta E=-\sum_{i=1,3}\delta t_i\sum_{s=1}^{S_{t_i}}\sum_{j,\eta}\sum_{i_n,\eta_n}\sum_{l_m,\tau_m}a_{i_1,\eta_1}(\bs{p}_1)a_{i_2,\eta_2}(\bs{p}_2)a^*_{l_1,\tau_1}(\bs{p}_1)a^*_{l_2,\tau_2}(\bs{p}_2)\Braket{(l_1,\tau_1),(l_2,\tau_2)|F\dagg_{j+r_{t_i}^{s,\eta},\eta+\eta_{t_i}}D\dagg_{j,\eta}D_{j+r_{t_i}^{s,\eta},\eta+\eta_{t_i}}F_{j,\eta}|(i_1,\eta_1),(i_2,\eta_2)}
\end{aligned}
\end{equation}
\end{widetext}
via the matrix elements of Eq.~\eqref{eq:me3}. Adapting the notation $i=(i,\eta)$ for clarity, $\Delta E$ then reduces to
\begin{equation} \label{eq:pt3}
\begin{aligned}
\Delta E&= -\delta t_i\sum_{s=1}^{S_{t_i}}\sum_{i_1,i_2}\biggl[a_{i_1}(\bs{p}_1)a_{i_2}(\bs{p}_2)a^*_{i_1+r^{s,i_1}_{t_i}}(\bs{p}_1)a^*_{i_2}(\bs{p}_2)-\\
&-a_{i_1}(\bs{p}_1)a_{i_2}(\bs{p}_2)a^*_{i_1+r^{s,i_1}_{t_i}}(\bs{p}_2)a^*_{i_2}(\bs{p}_1)\biggr]\,Q_c[i_1,i_2,i_1+r^{s,i_1}_{t_i}]+\\
&+\begin{Bmatrix}
i_1 \leftrightarrow i_2 \\
\bs{p}_1 \leftrightarrow \bs{p}_2 \\
\end{Bmatrix},
\end{aligned}
\end{equation}
which can be analyzed term-by-term. The second term in the square brackets of Eq.~\eqref{eq:pt3} involves a Fourier transform of $Q_c$ and vanishes in the limit $N\to \infty$. Inserting the known form of $a_{i}(\bs{p})$ into the first term of Eq.~\eqref{eq:pt3} and using the relation
\begin{equation} \label{eq:pt4}
\sum_{i_2}|C_{i_2}(\bs{p}_2)|^2Q_c[i_1,i_2,i_1+r^{s,i_1}_{t_i}]\xrightarrow[]{N \to \infty} Q_c[i_1,i_1+r^{s,i_1}_{t_i}],
\end{equation}
as well as $Q_c[i_1,i_1+r^{s,i_1}_{t_i}]=Q_c[(0,x),r^{1,x}_{t_i}]$ independent of $i_1$ and $s$ for $t_{i=1,3}$ due to symmetry, the final contribution to $\Delta E$ is $\varepsilon (\bs{p}_1)$ as defined in Eq.(\ref{eq:pert3}) from the main text. The last term in Eq.~\eqref{eq:pt3} which simply exchanges $i_1\leftrightarrow i_2$ and $\bs{p}_1\leftrightarrow \bs{p}_2$ then contributes $\varepsilon (\bs{p}_2)$. Again, this reasoning can be generalized in a straightforward way to arbitrary $N_f \ll N$.

\subsection{Dimer correlations} \label{dimercorrelation}
We show here that the fermionic anticommutation relations $\{f^\dagger_{\bs{p}_1},f^{\ }_{\bs{p}_2}\}=\delta_{\bs{p}_1,\bs{p}_2}$ hold exactly for momenta $\bs{p}_1$, $\bs{p}_2$ on the Brillouin zone (BZ) diagonal (i.e.~$p_x= p_y$), independent of system size. Consider the state
\begin{equation} \label{eq:dc1}
\{f\dagg_{\bs{p}_1},f_{\bs{p}_2}\}\Ket{0^*}=\sum_{i,\eta}a_{i,\eta}(\bs{p}_1)a^*_{i,\eta}(\bs{p}_2)\hat{N}_{i,\eta}\Ket{0^*}.
\end{equation}
If this state is to vanish for $\bs{p}_1\neq \bs{p}_2$, where we choose both $\bs{p}_1$ and $\bs{p}_2$ on the BZ diagonal, we need the prefactor of every basis configuration $\Ket{c}$ on the lattice to vanish. Here, $\Ket{c}$ represents a purely bosonic covering of the lattice as we apply $\{f^\dagger,f\}$ to the state $\Ket{0^*}$. Furthermore, we restrict to configurations $\Ket{c}$ of the topological sector with winding number zero. The corresponding prefactor is determined by the matrix element
\begin{equation} \label{eq:dc2}
d_c\equiv\Bra{c}\{f\dagg_{\bs{p}_1},f_{\bs{p}_2}\}\Ket{0^*}=\sum_{(i,\eta)\,\epsilon\,\{(j,\tau):\,\hat{N}_{j,\tau}\Ket{c}=\Ket{c}\}}a_{i,\eta}(\bs{p}_1)a^*_{i,\eta}(\bs{p}_2),
\end{equation}
where the sum runs over all links that are occupied by a dimer in the configuration $\Ket{c}$. Assume now we have some configuration $\Ket{c_0}$ where the value $d_{c_0}$ has already been calculated. Upon flipping a plaquette $(l,\nu),(l+\hat{\bar{\nu}},\nu)$ in $\Ket{c_0}$ we obtain a new state $\Ket{c_1}$ with corresponding prefactor
\begin{equation} \label{eq:dc3}
\begin{aligned}
\Delta d=d_{c_1}-d_{c_0}=a_{l,\bar{\nu}}(\bs{p}_1)a^*_{l,\bar{\nu}}(\bs{p}_2)+a_{l+\hat{\nu},\bar{\nu}}(\bs{p}_1)a^*_{l+\hat{\nu},\bar{\nu}}(\bs{p}_2)-\\
-a_{l,\nu}(\bs{p}_1)a^*_{l,\nu}(\bs{p}_2)-a_{l+\hat{\bar{\nu}},\nu}(\bs{p}_1)a^*_{l+\hat{\bar{\nu}},\nu}(\bs{p}_2).
\end{aligned}
\end{equation}
One can then compute the prefactors of all configurations within a fixed winding number sector by starting from a given configuration $\Ket{c_0}$ and subsequently going through all possible plaquette flips on the lattice. This reasoning corresponds to going the Ergodic theorem "backwards", i.e. instead of computing the desired correlation function as an average over an ensemble, we introduce a dynamics in the form of plaquette flips and compute the correlation of interest along the course of the plaquette flips.

Demanding that the prefactors $d_{c}$ vanish for all configurations in a winding number sectors then amounts to finding a starting configuration $\Ket{c_0}$ for which $d_{c_0}=0$ and showing that $\Delta d$ from Eq.~\eqref{eq:dc3} vanishes. For the latter, we plug the known form of $a_{i,\eta}(\bs{p})$ into Eq.~\eqref{eq:dc3} to obtain
\begin{equation} \label{eq:dc4}
\begin{aligned}
&\Delta d=e^{i(\bs{p}_1-\bs{p}_2)\cdot l}\bigl[C_{\bar{\nu}}(\bs{p}_1)C^*_{\bar{\nu}}(\bs{p}_2)+e^{i(\bs{p}_1-\bs{p}_2)\cdot \hat{\nu}}C_{\bar{\nu}}(\bs{p}_1)C^*_{\bar{\nu}}(\bs{p}_2)\\
&\qquad\qquad\qquad-C_{\nu}(\bs{p}_1)C^*_{\nu}(\bs{p}_2)-e^{i(\bs{p}_1-\bs{p}_2)\cdot \hat{\bar{\nu}}}C_{\nu}(\bs{p}_1)C^*_{\nu}(\bs{p}_2)\bigr]\\
&\propto (1+e^{ip_{1,x}})(1+e^{-ip_{2,x}})+e^{i(\bs{p}_1-\bs{p}_2)\cdot \hat{y}} (1+e^{ip_{1,x}})(1+e^{-ip_{2,x}})\\
&\quad- (1+e^{ip_{1,y}})(1+e^{-ip_{2,y}})-e^{i(\bs{p}_1-\bs{p}_2)\cdot \hat{x}} (1+e^{ip_{1,y}})(1+e^{-ip_{2,y}})
\end{aligned}
\end{equation}
which reduces to zero for $p_{1/2,x}=p_{1/2,y}$. For a starting configuration $\Ket{c_0}$ which corresponds to a columnar VBS state, which is part of the possible zero winding number configurations on the square lattice, we can easily calculate $d_{c_0}=0$ for $\bs{p}_1\neq \bs{p}_2$ on the diagonal. Hence, for two momenta $\bs{p}_1,\bs{p}_2$ on the BZ diagonal, $\{f\dagg_{\bs{p}_1},f_{\bs{p}_2}\}=\delta_{\bs{p}_1,\bs{p}_2}$ holds exactly; for every other pair of momenta it is valid in the limit considered before.

We can relate the relation we have just shown with the Fourier transform of the classical dimer correlation function to obtain an exact result. For $\bs{p}_1,\bs{p}_2$ on the BZ diagonal we can write
\begin{equation} \label{eq:dc5}
\begin{aligned}
\delta_{\bs{p}_1,\bs{p}_2}&=\sum_{c}|d_{c}|^2=\Vert\{f\dagg_{\bs{p}_1},f_{\bs{p}_2}\}\Ket{0^*}\Vert^2=\\
&=\frac{4}{N^2}\sum_{i,\eta,j,\tau}e^{i(\bs{p}_1-\bs{p}_2)\cdot (i-j)}Q_c[(i,\eta),(j,\tau)].
\end{aligned}
\end{equation}
We have thus found an exact sum rule for the Fourier transform of the classical dimer correlations on the BZ diagonal.

\subsection{The RK wave function} \label{rkwavefunction}
As mentioned in the main text, our ground state construction reproduces the unique RK wave function when the fermion in the $N_f=1$ ground states $\Ket{\bs{p}}$ is replaced by a bosonic dimer. We show now that all states with $\bs{p}\neq0$ vanish in the purely bosonic case, by proving that their overlap with all basis states vanishes. This can be shown by a simple construction similar to the one employed in the previous section: take $\Ket{\bs{p}}=\sum_{i,\eta}a_{i,\eta}(\bs{p})\Ket{i,\eta}$, with a boson at the link $(i,\eta)$, and let $\Ket{c}$ be an arbitrary (zero winding number) purely bosonic configuration on the square lattice. From the overlap 
\begin{equation}
\Braket{c|\bs{p}}=\sum_{(i,\eta): \hat{N}_{i,\eta}\Ket{c}=\Ket{c}}a_{i,\eta}(\bs{p})=:b_c,
 \end{equation}
we see that upon flipping a plaquette $(j,\nu),(j+\hat{\bar{\nu}},\nu)$ in $\Ket{c}$, the change in $b_c$ is given by 
\begin{equation}
\Delta b_c=a_{j,\bar{\nu}}(\bs{p})-a_{j,\nu}(\bs{p})+a_{j+\hat{\nu},\bar{\nu}}(\bs{p})-a_{j+\hat{\bar{\nu}},\nu}(\bs{p})=0 
\end{equation}
and vanishes according to the defining Eq.(\ref{eq:egs10}) for the $a_{i,\eta}(\bs{p})$ from the main text. For an arbitrary starting configuration $\Ket{c_0}$ (choose e.g. again the columnar state) we can check that $b_{c_0}=0$ for every $\bs{p}\neq 0$. Thus, since $\bs{p}\neq 0$ yields vanishing overlap with every basis state of the associated Hilbert space, the only remaining ground state is the $\bs{p}=0$ RK-state as required.


\end{document}